\documentclass[journal]{IEEEtran}
\usepackage[utf8]{inputenc}
\usepackage{amsmath}
\usepackage{listings}
\usepackage{cite}
\usepackage{adjustbox}
\usepackage{xcolor}
\usepackage{color,soul}
\usepackage{multirow}
\usepackage[
  separate-uncertainty = true,
  multi-part-units = repeat
]{siunitx}
\usepackage{algorithm}
\usepackage{graphicx}
\usepackage{textcomp}
\usepackage{hyperref}
\usepackage{caption}
\usepackage{subcaption}
\usepackage{authblk}
\usepackage{caption}
\usepackage{amsfonts}
\usepackage[justification=centering]{caption}

\begin{document}

\date{11 March 2022}

\title{RES-HD: Resilient Intelligent Fault Diagnosis Against Adversarial Attacks Using Hyper-Dimensional Computing}

\author{Onat Gungor, Tajana S. Rosing, \IEEEmembership{Fellow, IEEE}, and Baris Aksanli, \IEEEmembership{Member, IEEE}
\thanks{Onat Gungor is with the Electrical and Computer Engineering Department, University of California San Diego, La Jolla, CA 92093 USA and with the Electrical and Computer Engineering Department, San Diego State University, San Diego, CA 92182 USA (e-mail: ogungor@ucsd.edu).}
\thanks{Tajana Rosing is with the Computer Science and Engineering Department, University of California San Diego, La Jolla, CA 92093 USA (e-mail: tajana@ucsd.edu).}
\thanks{Baris Aksanli is with the Electrical and Computer Engineering Department, San Diego State University, San Diego, CA 92182 USA (e-mail: baksanli@sdsu.edu).}}


\maketitle
\pagestyle{plain}
\pagenumbering{gobble}

\begin{abstract}
Industrial Internet of Things (I-IoT) enables fully automated production systems by continuously monitoring devices and analyzing collected data. Machine learning methods are commonly utilized for data analytics in such systems. Cyberattacks are a grave threat to I-IoT as they can manipulate legitimate inputs, corrupting ML predictions and causing disruptions in the production systems. Hyper-dimensional computing (HDC) is a brain-inspired machine learning method that has been shown to be sufficiently accurate while being extremely robust, fast, and energy-efficient. In this work, we use HDC for intelligent fault diagnosis against different adversarial attacks.  Our black-box adversarial attacks first train a substitute model and create perturbed test instances using this trained model. These examples are then transferred to the target models. The change in the classification accuracy is measured as the difference before and after the attacks. This change measures the resiliency of a learning method. Our experiments show that HDC leads to a more resilient and lightweight learning solution than the state-of-the-art deep learning methods. HDC has up to 67.5\% higher resiliency compared to the state-of-the-art methods while being up to 25.1$\times$ faster to train. 
\end{abstract}

\begin{IEEEkeywords}
Industrial internet of things (I-IoT), predictive maintenance, intelligent fault diagnosis, hyper-dimensional computing, adversarial attacks, resilient machine learning 
\end{IEEEkeywords}

\section{Introduction}
\IEEEPARstart{I}{ndustry} 4.0 revolutionized monitoring, analysis, and automation of production systems through smart technology \cite{industry4.0}. It is mainly powered by the Industrial Internet of Things (I-IoT). I-IoT is the interconnection of smart devices that enables full automation, remote monitoring, and predictive maintenance. However, small-scale I-IoT devices with their limited computation and communication capabilities make I-IoT system an easy target for possible cyberattacks. System vulnerabilities (e.g., network protocols, insecure data transfer and storage) can be discovered by an attacker, and used to sabotage communication, do physical damage, alter existing data, or prevent asset availability \cite{tuptuk2018security}. The average estimated losses were \$10.7 million per breach of data among manufacturing organizations in Asia Pacific in 2019 \cite{microsoft2019}.
To minimize these costs, cyber-security measures should be taken such as cyber-security awareness training, keeping software  updated, installing firewalls, using strong passwords and others \cite{he2019improving}. 


Abundant system monitoring data in I-IoT systems makes data-driven predictive maintenance (PDM) popular.  Machine learning (ML) methods are commonly used for identifying best maintenance schedules \cite{gungor2021opelrul}. Relationship between past data and failure characteristics of a piece of equipment have been studied extensively \cite{wang2021bearing, lei2019fault, tao2019spur, gungor2021enfes}. Intelligent fault diagnosis (IFD) is a key data-driven PDM application that finds and classifies different fault types before they occur. 
The success of these ML-based methods heavily depends on input data. Adversarial attacks against ML methods manipulate legitimate inputs and force the trained model to produce incorrect outputs leading to incorrect predictions. Since ML is in the center of intelligent fault diagnosis, these attacks may have serious consequences such as undetected failures \cite{mode2020crafting}. There is a need for novel intelligent learning solutions that can stay resilient against various adversarial attacks.

Hyper-dimensional computing (HDC) was introduced as a brain-inspired learning solution for robust and efficient learning \cite{imani2017exploring}. HDC encodes raw data into high-dimensional vectors (i.e., hypervectors) and then performs simple operations in this space, e.g., element-wise addition, dot product. 
HDC has been employed in a range of applications such as activity recognition \cite{rasanen2015sequence}, speech recognition \cite{imani2017voicehd}, and biomedical signal processing \cite{moin2021wearable}. 
To the best of our knowledge, it has not been used for intelligent fault diagnosis.  

In this work, we propose HDC as a resilient learning solution against different black-box adversarial attacks for intelligent fault diagnosis (IFD). Our black-box attack is based on a transferable attack strategy \cite{bhambri2019survey}. 
We first train a substitute deep learning model, a wide deep convolutional neural network (WDCNN), and create artificial test samples using this trained model. We then transfer these instances to the target methods (e.g., LSTM, GRU, HDC). In testing, we measure the classification accuracy of the target models before and after the attacks. The accuracy change gives us the resiliency of a target method. We show that HDC is the most resilient and lightweight method, outperforming all deep learning (DL) methods on commonly used CWRU Bearing dataset \cite{CWRUDataset}. HDC is up to 67.5\% more resilient and 25.1$\times$ faster during training compared to the state-of-the-art deep learning methods.

\section{Related Work}
Industrial Internet of Things (I-IoT) is an adaptation of traditional IoT for production environments focusing on machine-to-machine communication, big data, and machine learning for higher system efficiency and reliability \cite{xu2018survey}. I-IoT systems are often insufficiently secure and vulnerable due to off-the-shelf communication protocols \cite{tuptuk2018security, wu2018cybersecurity, enisa2018}. Furthermore, there are many assets that are vulnerable to cyber-attacks such as operating systems, application software, communication protocols, and smart devices \cite{wu2018cybersecurity}. An adversary, attacker, can exploit these vulnerabilities to arrange a cyberattack. There is an increasing cyberattack trend in recent years, for example, in 2019, cyberattacks against factories increased by more than 200\% \cite{cyber-attacks-smart-factory}. There are various cyberattacks against I-IoT systems such as denial of service, eavesdropping, side channel, and attacks against machine learning \cite{tuptuk2018security}. We focus on attacks against ML in this paper where an attacker corrupts the collected data or model parameters leading to worse prediction performance. These attacks are dangerous since data analytics is an indispensable part of I-IoT systems. They can result in serious outcomes, e.g., undetected failures in a system \cite{mode2020crafting}.     

Predictive maintenance determines an optimal maintenance schedule based on time-to-failure prediction of industrial assets \cite{gungor2021opelrul}. Data-driven predictive maintenance utilizes historical data to create ML models which are then used for fault diagnosis of industrial assets. Intelligent fault diagnosis (IFD) is a branch of data-driven PDM which discovers and classifies different fault types in advance. There are various IFD methods, such as convolutional neural network (CNN) \cite{wang2021bearing}, long short-term memory (LSTM) \cite{lei2019fault}, gated recurrent unit (GRU) \cite{tao2019spur}, ensemble learning \cite{gungor2021enfes}, etc. However, adversarial attacks against deployed ML models can lead to serious consequences for a PDM system such as delayed maintenance or replacement of a machine \cite{mode2020crafting}. Mode and Hoque \cite{mode2020crafting} analyze the impact of different adversarial attacks against ML for remaining useful life (RUL) prediction. They implement Fast Gradient Sign Method (FGSM) and Basic Iterative Method (BIM) to craft adversarial examples and compare different DL model performances under those attacks. They use the NASA C-MAPSS dataset \cite{saxena2008damage} and show that these attacks can cause up to 5$\times$ worse prediction performance. This work is limited in terms of the number of attack methods and DL models. Besides, they did not propose a novel methodology that can stay resilient against adversarial attacks. In our paper, we propose a novel learning solution which can stay resilient against different adversarial attacks. Our results are also more generalizable because we increase the number of attack scenarios and deep learning models significantly compared to the state-of-the-art.

Brain-inspired Hyperdimensional computing (HDC) was proposed as a light-weight computing method to perform cognitive tasks on devices with limited resources \cite{imani2017exploring}. HDC has three main stages \cite{morris2020multi}: 1) encoding: mapping data into high dimensional vectors, called hypervectors (HVs). 2) training: combining encoded HVs to create a model representing each class with a HV. 3) inference: comparing the test sample with the trained model to find the most similar class. The key property of HD is that it can provide strong robustness to noise \cite{morris2021hydrea} which might be really beneficial for I-IoT systems since collected sensor data contains noise in general \cite{liu2020noise}. There are many applications using HD such as multimodal sensor fusion \cite{rasanen2015sequence}, language recognition \cite{rahimi2016robust}, and speech recognition \cite{imani2017voicehd}. 
To the best of our knowledge, HD has not been used in PDM domain where it can provide both lightweight and robust learning solution. In our work, we propose HD as a resilient and efficient learning solution to adversarial attacks. Our experiments show that HD is more resilient against various adversarial attacks compared to the state-of-the-art DL methods while bringing significant computational advantage.

\section{Adversarial Attack Methods}
\label{adversarial-attacks}
Data-driven predictive maintenance (PDM) heavily uses deep learning models due to their great prediction performance \cite{zhang2019data}. Deep learning, as a branch of ML, has many security vulnerabilities such as program errors (e.g., memory depletion), and attacks at the time of its testing (e.g., worse prediction performance) \cite{tariq2020review}. Adversarial attacks against ML modify input data slightly in a way that is intended to cause an ML classifier to misclassify it. 

There are three types of adversarial attacks against ML in the literature: evasion, poisoning, and exploratory \cite{chakraborty2018adversarial}. Evasion attacks target compromising the test data, poisoning attacks contaminate the training data, and exploratory attacks gain knowledge about the learning algorithm without changing the data. We focus on evasion attacks since it is the most common type of attack in an adversarial setting \cite{chakraborty2018adversarial}. Evasion attacks can further be categorized into two groups: white-box and black-box attacks. While white-box attacks have detailed knowledge about the model (e.g., type of neural network along with the number of layers), algorithm used in training (e.g., gradient-descent optimization), the training data distribution, and the parameters of the fully trained model architecture, black-box attacks assume no knowledge about the underlying model. Hence, black-box attacks reflect more realistic scenarios in the sense that the attacker can be an outsider, with limited knowledge about the internal system details \cite{bhambri2019survey}. 
An adversary can craft a black-box attack using (1) a transferable attack strategy which trains a substitute model to emulate the original (target) model and (2) a query feedback mechanism that starts with random input to be sent to the target model. Based on the confidence score of inputs, attacker adds noise to the input under the acceptable perturbation error level.  

In this work, we adapt a transferable attack strategy where we select a single substitute model and craft examples using this model. We then transfer the created examples to target models. In this strategy, attacker does not need to know anything about the target models, yet have an access to input data. Adversary solely trains the substitute model and transfer created examples as the perturbed inputs start misclassified by the substitute model. 
We select 4 different state of the art adversarial attack methods using the gradient information of the loss function: fast gradient sign method (FGSM), basic iterative method (BIM), momentum iterative method (MIM), and robust optimization method (ROM). Our problem is intelligent fault diagnosis (IFD) where our models predict fault types before they actually happen. IFD is a time series classification problem since input is a time series sensor data, and output is a certain fault type (which is discrete). We define the following variables to explain different attack strategies:
\begin{itemize}
    \item $\theta$: parameters of the substitute model
    \item $x$: input data (collected sensor data)
    \item $y$: labels (fault types)
    \item $J(\theta, x, y)$: cost function to train the substitute model
\end{itemize}

\subsection{Fast Gradient Sign Method (FGSM)}
FGSM was proposed as an efficient adversarial attack method to fool the GoogLeNet model \cite{goodfellow2014explaining}. FGSM first calculates the gradient of the cost function with respect to the input of the neural network. Adversarial examples are then created based on the gradient direction:
\begin{equation}
    \ddot{x} = x + \epsilon * sign(\nabla_{x} J(\theta, x, y))
\end{equation}
where $\ddot{x}$ represents the crafted adversarial examples and $\epsilon$ denotes the amount of the perturbation. 

\subsection{Basic Iterative Method (BIM)}
BIM is an improved version of FGSM where FGSM is applied multiple times with really small step size \cite{kurakin2016adversarial}. BIM perturbs the original data in the direction of the gradient multiplied by the step size $\alpha$:
\begin{equation}
    \ddot{x} = x + \alpha * sign(\nabla_{x} J(\theta, x, y))
\end{equation}
where $\alpha$ is obtained by dividing the amount of perturbation ($\epsilon$) by the number of iterations ($I$): $\alpha = \epsilon / {I}$. Then, BIM clips the obtained time series elements to ensure that they are in the $\epsilon$-neighborhood of the original time series:  
\begin{equation}
    \ddot{x} = min\{x + \epsilon, max\{x - \epsilon, \ddot{x}\}\}
\end{equation}

\subsection{Momentum Iterative Method (MIM)}
Momentum Iterative Method (MIM) solves underfitting and overfitting problems in FGSM and BIM respectively by integrating momentum into the BIM \cite{dong2018boosting}. At each iteration i, the variable $g_{i}$ gathers the gradients with a decay factor $\mu$:
\begin{equation}
    g_{i+1} = \mu * g_{i} + \dfrac{\nabla_{x} J(\theta, \ddot{x}_{i}, y)}{\|\nabla_{x} J(\theta, \ddot{x}_{i}, y)\|_{1}}
\end{equation}
where the gradient is normalized by its $L_{1}$ distance. The perturbed data for the next iteration is created in the direction of the sign of $g_{i+1}$ with a step size $\alpha$:
\begin{equation}
    \ddot{x}_{i+1} = \ddot{x}_{i} + \alpha * sign(g_{i+1})
\end{equation}
In MIM, the algorithm also ensures that the crafted adversarial examples $\ddot{x}$ satisfy the $L_{\infty}$ norm bound constraint, i.e. $\|\ddot{x} - x\|_{\infty} \leq \epsilon$. Besides, if $\mu = 0$, MIM boils down to BIM. 

\subsection{Robust Optimization Method (ROM)}
The goal of a supervised learning problem is to find model parameters $\theta$ that minimize the empirical risk $\mathbb{E}_{(x, y) \sim \Xi}[J(\theta, x, y)]$ where $\Xi$ is the underlying supervised data distribution. However, this formulation cannot handle the change in input data. To solve this problem, set of allowed perturbations $\Delta$ is introduced. 
Then, empirical risk formulation is modified by feeding samples from the distribution $\Xi$ directly into the loss function $J$ leading to the following min-max (saddle point) optimization formulation \cite{madry2017towards}:
\begin{equation}
    \underset{\theta}{\text{min}} \ \zeta(\theta), \ where \ \zeta(\theta) = \mathbb{E}_{(x, y) \sim \Xi}[\underset{\delta \in \Delta}{\text{max}} \ J(\theta, x + \delta, y)]
\end{equation}
Here, while inner maximization finds an adversarial version of a given data point $x$ that achieves a high loss, outer minimization discovers model parameters to minimize the adversarial loss given by the inner attack problem. To solve the robust optimization problem, ROM replaces every instance with its FGSM-perturbed counterpart. 

The selected methods utilize loss gradient information to craft adversarial examples by adding different amounts of perturbation. In that sense, they represent different attack scenarios. An adversary, who is able to access the I-IoT system data (e.g., collected sensor measurements), can implement these methods using a substitute model and harm the prediction performance of target models without being detected. 

\section{Proposed Framework}
\label{framework}
In this work, we use hyper-dimensional computing (HDC) for intelligent fault diagnosis and measure the resiliency of this method against different adversarial attacks.  To the best of our knowledge, HDC has not been used in this context before. 
Deep learning models, on the other hand, have been used extensively for I-IoT applications\cite{zhang2019data}. However, these models are vulnerable to adversarial attacks which can deteriorate the prediction performance. HDCs resilience makes it a great alternative.  
\begin{figure}[h]
\centering\includegraphics[width=0.95\linewidth]{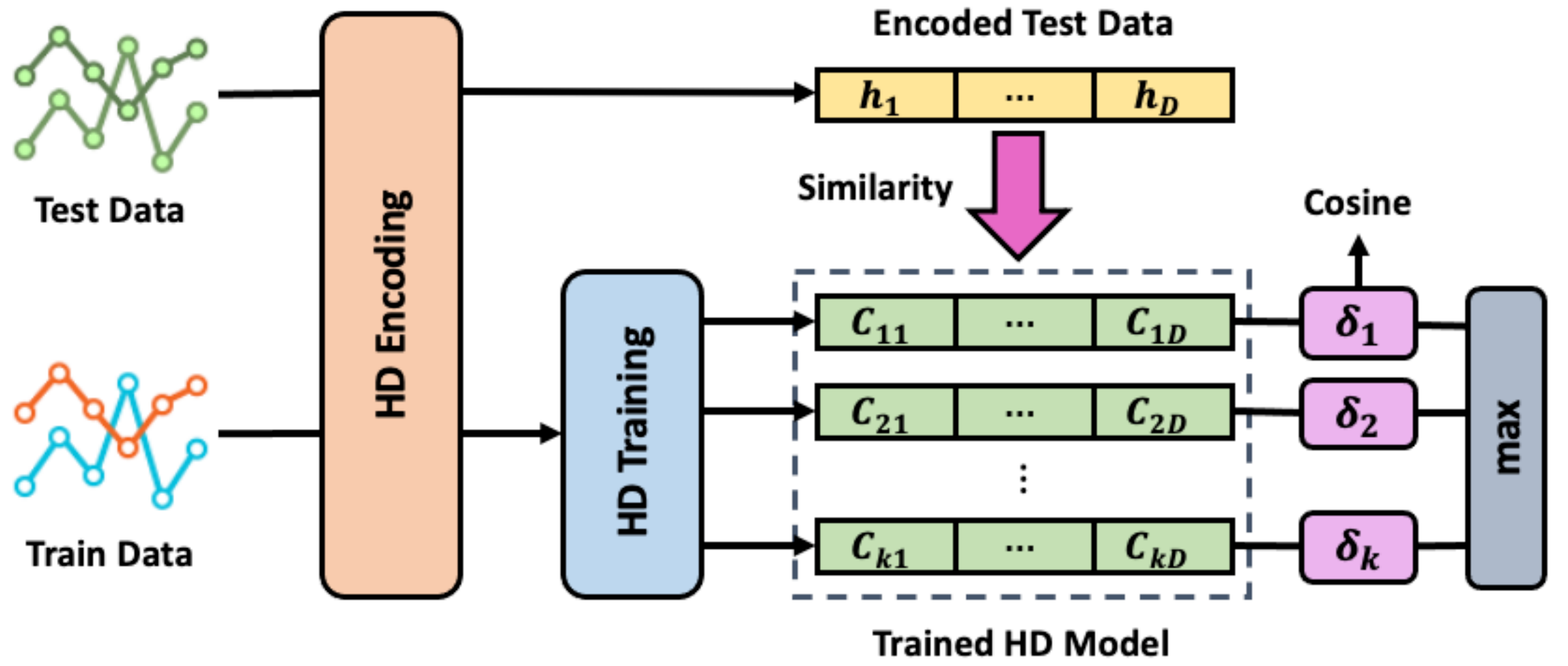}
\caption{HDC Learning Framework}
\label{hd-framework}
\end{figure}

\begin{figure*}[]
\centering\includegraphics[width=0.87\linewidth]{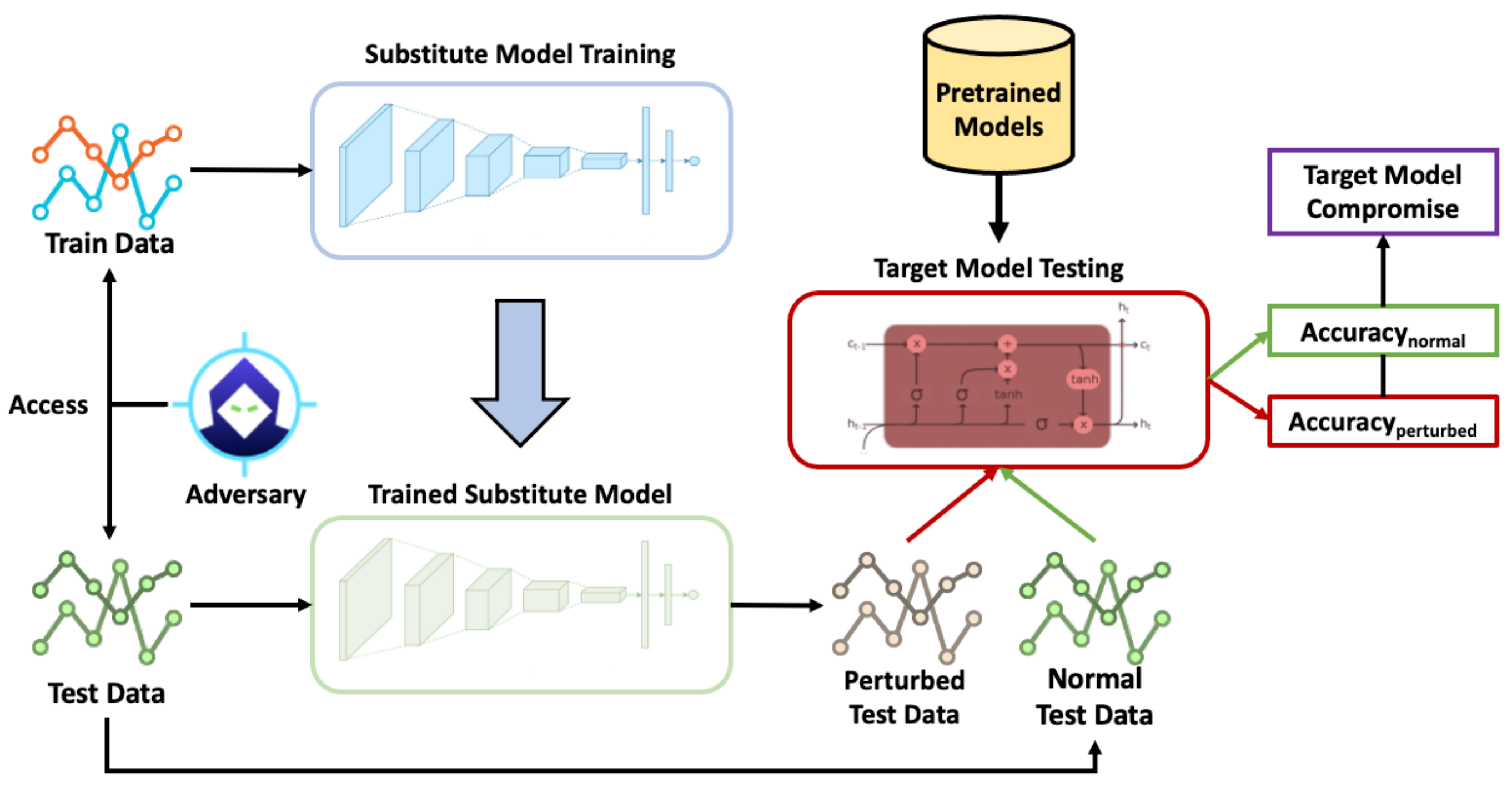}
\caption{Black-box Attack Framework}
\label{black-box-framework}
\end{figure*}

\subsection{Hyper-dimensional Computing (HDC)}
HDC is an emerging computing paradigm inspired by the fact that brains take sensing data and map it into a high dimensional sparse representation before analysis \cite{babadi2014sparseness}. HDC mimics this by mapping each data point into high-dimensional space $\mathcal{D}$. All computational tasks are performed in $\mathcal{D}$-space using simple operations such as element-wise additions and dot products \cite{thomas2020theoretical}. HDC is especially suitable for several learning tasks in IoT systems as (1) it is computationally efficient (2) it supports single-pass training (no back-propagation needed), and (3) it provides strong robustness to noise \cite{zou2021spiking}. Fig. \ref{hd-framework} illustrates HDC learning framework which contains 3 main stages: encoding, training, and inference. 

\subsubsection{Encoding}
The first step of HDC is to encode input data into hyper-vectors. Most of the proposed encoding methods \cite{rahimi2020hyperdimensional, nazemi2020synergiclearning} linearly combine the hyper-vectors corresponding to each feature, resulting in sub-optimal classification quality \cite{zou2021spiking}. In this work, we use non-linear encoding which considers the non-linear interactions between the feature values with different weights and exploits the kernel trick. This encoding approach is based on a study which shows that the Gaussian kernel function can be approximated by the dot product of two vectors \cite{rahimi2007random}. Assume that an input vector in original space $\Vec{\Xi} = \{\xi_{1}, \xi_{2}, \dots \xi_{n}\} \in \mathbb{R}^n$. Encoded high-dimensional vector is represented as $\Vec{\mathcal{H}} = \{h_{1}, h_{2}, \dots, h_{\mathcal{D}} \} \in \mathbb{R}^\mathcal{D}$ where $\mathcal{D} \gg n$. Encoding from $\Vec{\Xi}$ to $h_{i}$ is based on the following equation:
\begin{equation}
    h_{i} = cos(\Vec{\Xi} \cdot \Vec{\mathcal{B}_{i}} + b_{i}) sin(\Vec{\Xi} \cdot \Vec{\mathcal{B}_{i}})
\end{equation} 

where $\Vec{\mathcal{B}_{k}}$s are randomly chosen of dimension $\mathcal{D} \simeq 10k$ and $b_{i} \sim \mathcal{U}(0, 2\pi)$. That is, $\Vec{\mathcal{B}}_{kj} \sim \mathcal{N}(0,1)$ and $\delta(\Vec{\mathcal{B}}_{k1}, \Vec{\mathcal{B}}_{k2}) \simeq 0$, where $\delta$ is the cosine similarity.  

\subsubsection{Training}
The second step of HDC, model training consists of two steps to generate hyper-vectors representing each class. The first step, \textbf{initial training}, performs element-wise addition of all encoded hyper-vectors in each existing class. Let's assume that $\Vec{\mathcal{H}}_{i}$ is the encoded hyper-vector of input $i$. We know that each input $i$ belongs to a class $j$. Hence, $\Vec{\mathcal{H}}_{i}^{j}$ denotes the hyper-vector for input $i$ from class $j$. In the initial training, HDC simply adds all hyper-vectors of the same class to generate the final model hyper-vector:
\begin{equation}
    \Vec{C}^{j} = \eta \Vec{\mathcal{H}}_{0}^{j}+ \eta \Vec{{H}}_{1}^{j}+\dots = \sum_{m} \eta \Vec{\mathcal{H}}_{m}^{j}
\end{equation}

where $\eta$ is the learning rate. This process is also called as one-pass training since each input is visited only once and added to class hyper-vectors. The second step of HDC training, \textbf{retraining}, performs model adjustment by iteratively going through the training dataset. Retraining is beneficial for HDC to improve the prediction accuracy. During this step, the encoded hyper-vector of each input is created again, and its similarity with the existing class hyper-vectors is checked. If HDC misclassifies, say that $\Vec{\mathcal{H}}^{j}$ from class $\Vec{C}^{j}$ is predicted as class $\Vec{C}^{k}$, it updates its model as follows:
\begin{equation}
\begin{aligned}
    \Vec{C}^{j} = \Vec{C}^{j} + \eta \Vec{\mathcal{H}}^{j} \\
    \Vec{C}^{k} = \Vec{C}^{k} - \eta \Vec{\mathcal{H}}^{j}
\end{aligned}
\end{equation}
which means that the information of $\Vec{\mathcal{H}}^{j}$ causing misclassification to $\Vec{C}^{k}$ is discarded.

\subsubsection{Inference}
In the last step, HDC checks the similarity of each encoded test data with the class hyper-vector. Most commonly, cosine similarity is used for the similarity check although other metrics (e.g., Hamming distance) could be suitable based on the problem. To calculate cosine similarity between hyper-vector $\Vec{\mathcal{H}}$ and class hyper-vector $\Vec{C}^{j}$:   
\begin{equation}
    cos(\Vec{\mathcal{H}}, \Vec{C}^{j}) = \dfrac{\Vec{\mathcal{H}} \cdot \Vec{C}^{j}}{\|\Vec{\mathcal{H}}\| \cdot \|\Vec{C}^{j}\|}
\end{equation}
which is calculated by the dot product of the $\Vec{\mathcal{H}}$ and $\Vec{C}^{j}$ divided by the product of these two vectors' lengths. As an output of this step, HDC provides the most similar class.   


 

\subsection{Black-box Attack Framework}
In this work, we use a black-box transferable attack strategy which first trains a substitute model and crafts new test instances using the trained substitute model. For full list of DL methods used in this paper, you can refer to the Section \ref{exp-setup}. As our substitute model, we select wide deep convolutional neural network (WDCNN) since it is one of the most commonly used DL methods in intelligent fault diagnosis \cite{zhang2019limited, kolar2020fault, gungor2021enfes}. For our black-box attack setting, we assume that an adversary has access to the training and test data, yet does not know anything about the attacked (target) models. We illustrate our black-box attack framework in Fig. \ref{black-box-framework}. Attacker first trains the substitute model (WDCNN) and use the trained model to create perturbed test data. Attacker can employ different attack strategies to obtain perturbed test data (see Section \ref{adversarial-attacks} for the attack strategies used in this paper). Afterwards, adversary sends these crafted examples to the target models in testing time. In Fig. \ref{black-box-framework}, we give long short-term memory (LSTM) as the target model for illustration purposes. However, note that there is a pool of pretrained target models (LSTM, GRU, CRNN, CLSTM, CGRU, SCRNN, SCLSTM, SCGRU, HDC) adversary is not aware of (thus black-box attack). Attacker simply sends the created examples to the target models to see if the attack will be successful or not. We measure the attack success based on change in test data classification accuracy before and after the attack where accuracy is defined as: 
\begin{equation}
Accuracy = \dfrac{\text{Number of correct predictions}}{\text{Total number of test samples}}
\end{equation}
Change in accuracy gives us the resiliency of a learning method which we measure by the metric called \textit{Compromise} which is formulated as: 
\begin{equation}
Compromise = \dfrac{Accuracy_{normal}}{Accuracy_{perturbed}} 
\end{equation}
where $Compromise > 1$ (under the assumption that attacks lead to worse prediction performance). The \textbf{smaller} the compromise value is, the \textbf{more resilient} the model becomes against the adversarial attack. For instance, given two methods CRNN and LSTM, and their compromise values 5 and 2 respectively, we can conclude that LSTM is \textbf{more resilient} against the adversarial attack. If we have $M$ number of adversarial attacks ($M > 1$), then we need to calculate the mean compromise value for each learning method as follows:
\begin{equation}
Compromise_{mean} = \left(\sum_{i=1}^{M} \dfrac{Accuracy_{normal}}{Accuracy_{perturbed}^{i}}\right) / M
\label{mean-compromise}
\end{equation}
Because we have multiple attack strategies (where $M = 4$), mean compromise gives a more accurate idea about single model resiliency. Overall, we obtain mean compromise value for each learning method and use this metric for our experimental analysis. Furthermore, to show the HDC resiliency improvement, we define the following improvement metric:
\begin{equation}
    Improvement = \left(\dfrac{Compromise_{DL} - Compromise_{HDC}}{Compromise_{DL}}\right)
    \label{eq-improvement}
\end{equation}
where $Compromise_{DL}$ denotes the single DL model mean compromise value, and $Compromise_{HDC}$ is the HDC mean compromise. We report the improvement in percentage (\%). Improvement signifies the resiliency of our HDC learner against adversarial attacks compared to a single DL model. 


\section{Experimental Analysis}
\label{experimental}
\subsection{Dataset Description}
\begin{figure}[]
\centering\includegraphics[width=0.95\linewidth]{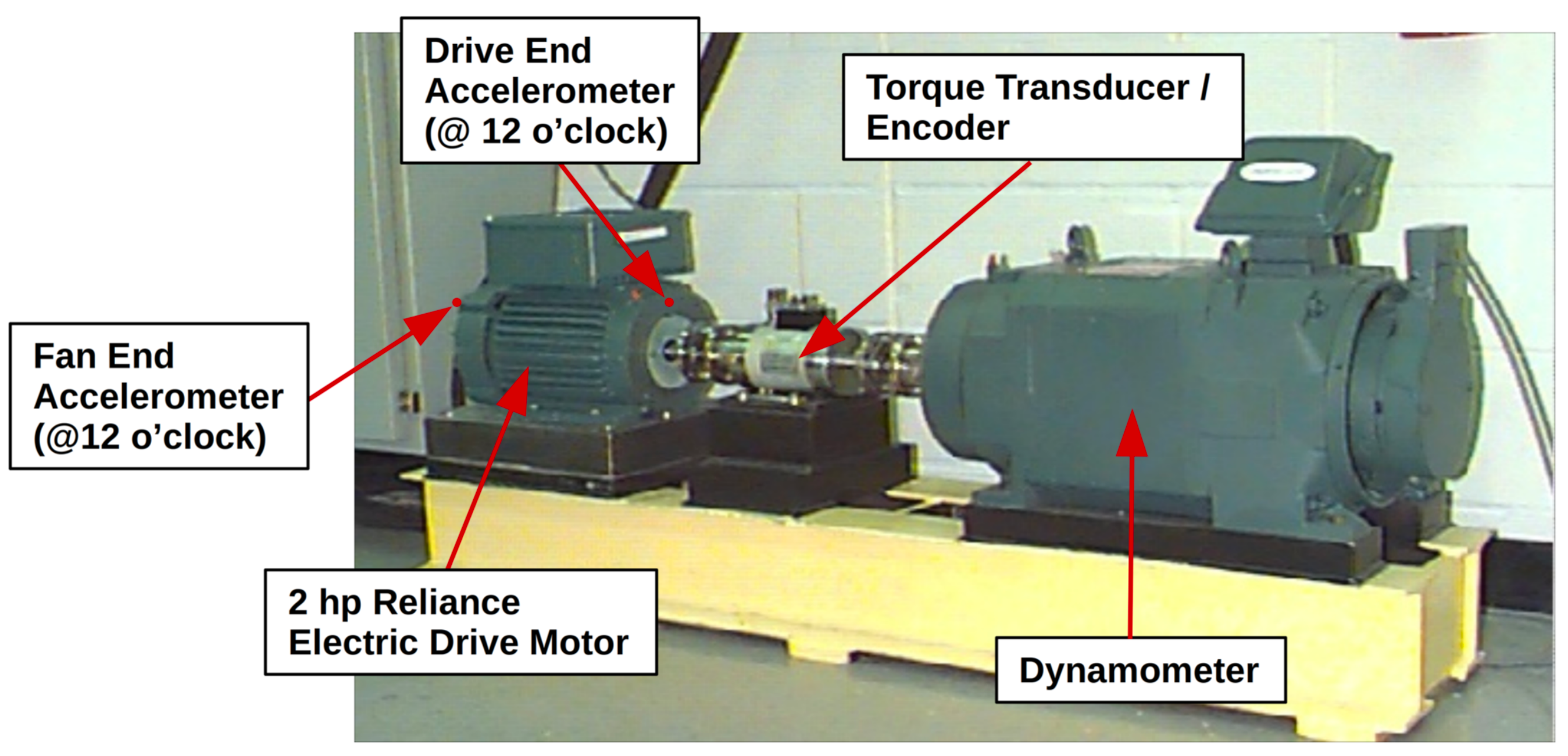}
\caption{CWRU Experimental Test Apparatus \cite{shenfield2020novel}}
\label{cwru-test}
\end{figure}

We use Case Western Reserve University (CWRU) Bearing dataset \cite{CWRUDataset}, a widely used benchmark for fault diagnosis. Rolling element bearing (REB) failure is one of the most frequent reasons for machine breakdown leading to severe loss of safety and property \cite{smith2015rolling}. Fig. \ref{cwru-test} represents the experimental test apparatus to collect this dataset. The data were collected from both the drive end accelerometer and the fan end accelerometer at 12k samples/second over a range of motor loads (from 0 hp to 3 hp). Both datasets (drive end and fan end) contain 19,800 training and 750 test samples. Bearing used in this experiment has three components: rolling element, inner race, and outer race. 9 different fault types are provided in the dataset based on the fault diameter (0.007, 0.014, and 0.021 inches) and the component (plus the normal bearing condition).

\subsection{Experimental Setup}
\label{exp-setup}
\textbf{Selected Deep Learning (DL) Methods: } We select 9 deep learning methods from recurrent (LSTM, GRU), convolutional (WDCNN), and hybrid (CRNN, SCRNN) architectures. We cover a good range of DL methods from different architectures, increasing the generalizability of our study. 
\begin{itemize}
    \item \textbf{Long Short-Term Memory (LSTM) \cite{lei2019fault}:} Our network contains 2 LSTM layers having 64 and 32 units which are consecutively connected to fully connected feed forward neural network with 10 units representing each failure class (we have 10 failure types in total).
    \item \textbf{Gated Recurrent Unit (GRU) \cite{tao2019spur}:} We use the same network structure as LSTM by updating the LSTM layers to GRU.
    \item \textbf{Wide Deep Convolutional Neural Network (WDCNN) \cite{zhang2017new}}: WDCNN starts with a convolutional layer using kernel size of 64. It is connected to a pooling layer. This is then connected to multi stage convolutional layers with small kernels (kernel size of 3) and pooling layers. Whole CNN structure is finally connected to fully connected feed forward neural networks with 100 and 10 units. 
    \item \textbf{Convolutional Recurrent Neural Network (CRNN) \cite{shenfield2020novel}:} CRNN is a hybrid model which has two parallel paths: one path has WDCNN structure and another path has 1D wide convolutional layer (with a kernel size of 64) connected to single RNN layer with 128 units. These two layers are then concatenated and connected to feed forward neural network with 10 units. Here, for the RNN layer, we use RNN (CRNN), LSTM (CLSTM), and GRU (CGRU) models providing 3 different models. 
    \item \textbf{Simplified CRNN (SCRNN) \cite{shenfield2020novel}:} Here, we remove the WDCNN path from CRNN structure meaning that SCRNN only contains 1D wide convolutional layer (with a kernel size of 64) connected to single RNN layer with 128 units. Similarly, we use RNN (SCRNN), LSTM (SCLSTM), and GRU (SCGRU) models providing 3 different models.  
\end{itemize}

\textbf{Adversarial Attack Methods:} We select the following parameters for our adversarial methods: fast gradient sign, basic iterative, momentum iterative, and robust optimization \cite{fawaz2019adversarial, dong2018boosting, madry2017towards}: amount of perturbation ($\epsilon$): 0.1, step size ($\alpha$): 0.001, number of iterations ($I$): 100, decay factor ($\mu$): 1. 


\textbf{Parameter Selection:} For both deep learning methods and HDC, we use a sliding time window of size 100. We replicate each experiment 10 times and report average values where we run all experiments on a PC with 16 GB RAM and an 8-core 2.3 GHz Intel Core i9 processor. For \textbf{DL methods}, the following hyper-parameters are selected: \textit{Adam} optimizer with learning rate 0.001, \textit{relu} activation function, batch size of 16, and a max number of epochs of 100 where callback is activated (patience is set to 10 for validation data). For \textbf{HDC}, we select the following parameters: encoding: non-linear, hyper-vector dimension size: 10,000, learning rate: 0.005, number of epochs: 100, similarity metric: cosine.

\textbf{Number of Training Samples:} We measure the selected methods' resiliency by using different number of training samples while using the whole test data. 
Specifically, our smallest experiment configuration uses 1.2\% (240 samples) of the whole training data where we double this ratio until we reach approximately 38.8\% (7680 samples). We call this ratio sample training ratio (STR) for the rest of this section. Considering different STRs is crucial for intelligent fault diagnosis (IFD) since it might not always be feasible to label fault data for the whole training dataset. IFD methods should perform well under limited supervision \cite{gungor2021enfes}.  

\begin{figure}[]
\centering
\begin{subfigure}{.87\linewidth}
    \centering
    \includegraphics[width=\linewidth]{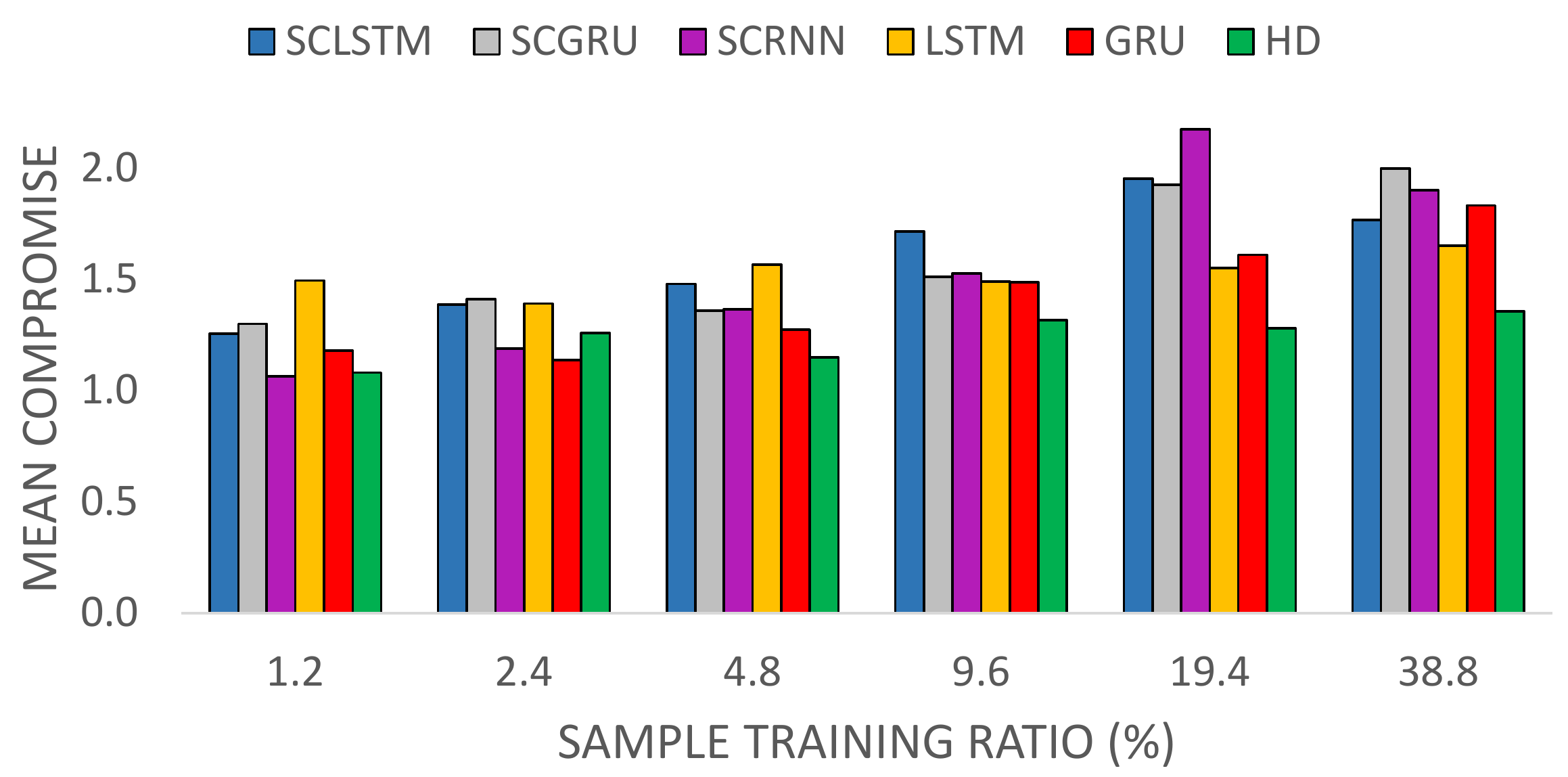}
    \caption{\textbf{Drive End Dataset}}
    \label{fig:drive-end-compromise}
\end{subfigure}
\begin{subfigure}{.87\linewidth}
    \centering
    \includegraphics[width=\linewidth]{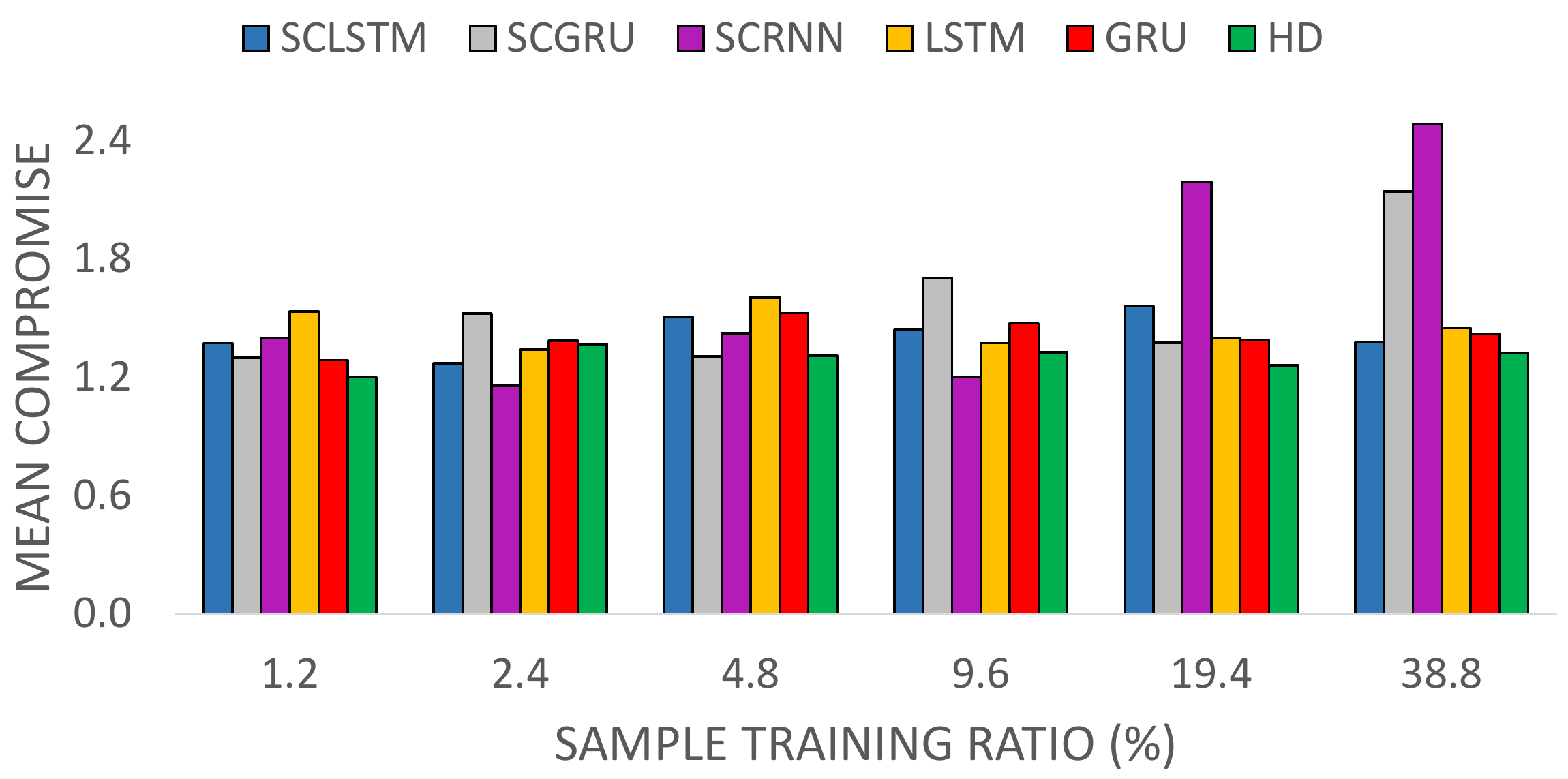}
    \caption{\textbf{Fan End Dataset}}
    \label{fig:fan-end-compromise}
\end{subfigure}
\newline
\caption{Mean Compromise Analysis}
\label{resiliency-analysis}
\end{figure}

\subsection{Resiliency Analysis}

\textbf{Mean Compromise Comparison:} We analyze the resiliency of selected DL models and HDC using mean compromise metric defined in Equation \ref{mean-compromise}. Fig. \ref{resiliency-analysis} shows the mean compromise values of the 6 most resilient learning methods under different sample training ratios for drive end (Fig. \ref{fig:drive-end-compromise}) and fan end (Fig. \ref{fig:fan-end-compromise}) datasets. In these figures, x-axis represents the selected STRs and y-axis gives the mean compromise values where each color represents different learning method. Based on these figures, we can observe that the mean compromise of a DL method changes significantly. To illustrate, while SCRNN (represented with purple color) is the most resilient method for really small STRs (e.g., 1.2, 2.4), it becomes the least resilient algorithm as we reach the maximum STR. For drive end dataset, we can observe that HDC is the most resilient method for STRs greater than 2.4\%. As the number of training samples increases, HD becomes a more resilient method compared to the DL algorithms. We can make a similar observation for fan end dataset as well. For STRs larger than 9.6\%, HD is the most resilient method against adversarial attacks. To present a single mean compromise value (for better understanding), we calculate the average of mean compromise values over all STRs. Table \ref{mean-compromise-comparison} presents these average compromise values for all the methods. When we compare DL methods (i.e., all methods excluding HD), we can observe that recurrent neural network structures are the most resilient. Specifically, GRU is the most resilient DL method with an average compromise value of 1.42 for both datasets. This observation can be attributed to the fact that our hybrid DL model structures contain convolutional layers. Since our crafted examples are based on wide deep convolutional neural network, more test examples are able to deceive hybrid methods. Most importantly, according to Table \ref{mean-compromise-comparison}, \textbf{HD} is the \textbf{most resilient} method on average outperforminng other DL methods at both datasets. This shows that HD provides a resilient learning solution performing well even under different black-box adversarial attack configurations.  

\begin{table}[]
\centering
\caption{Average Compromise Comparison}
\scalebox{0.95}{
\begin{tabular}{|c|c|c|}
\hline
\textbf{Method / Dataset} & \textbf{Drive end} & \textbf{Fan end} \\ \hline
\textbf{CGRU}             & 2.77               & 2.52             \\ \hline
\textbf{CLSTM}            & 2.52               & 2.69             \\ \hline
\textbf{CRNN}             & 2.30               & 2.77             \\ \hline
\textbf{SCLSTM}           & 1.59               & 1.42             \\ \hline
\textbf{SCGRU}            & 1.58               & 1.56             \\ \hline
\textbf{SCRNN}            & 1.54               & 1.69             \\ \hline
\textbf{LSTM}             & 1.52               & 1.45             \\ \hline
\textbf{GRU}              & 1.42               & 1.42             \\ \hline
\textbf{HD}               & \textbf{1.24}      & \textbf{1.30}    \\ \hline
\end{tabular}}
\label{mean-compromise-comparison}
\end{table}

\begin{table}[t]
\centering
\caption{Average and Maximum Resiliency Improvement of HDC over DL Methods}
\scalebox{0.8}{
\begin{tabular}{|c|cc|cc|}
\hline
                   & \multicolumn{2}{c|}{\textbf{Drive end}}                            & \multicolumn{2}{c|}{\textbf{Fan end}}                              \\ \hline
\textbf{DL Method} & \multicolumn{1}{c|}{\textbf{Average (\%)}} & \textbf{Maximum (\%)} & \multicolumn{1}{c|}{\textbf{Average (\%)}} & \textbf{Maximum (\%)} \\ \hline
\textbf{CGRU}    & \multicolumn{1}{c|}{\textbf{51.34}}         & \textbf{61.9}                  & \multicolumn{1}{c|}{44.25}                   & 64.5                  \\ \hline
\textbf{CLSTM}    & \multicolumn{1}{c|}{48.1}         & 54.1                  & \multicolumn{1}{c|}{48.3}                   & \textbf{67.5}                   \\ \hline
\textbf{CRNN}    & \multicolumn{1}{c|}{42.1}         & 52.4                  & \multicolumn{1}{c|}{\textbf{49.9}}                   & 64.4                  \\ \hline
\textbf{SCLSTM}    & \multicolumn{1}{c|}{21.1}         & 34.5                  & \multicolumn{1}{c|}{8.1}                   & 19.3                  \\ \hline
\textbf{SCGRU}     & \multicolumn{1}{c|}{20.2}                  & 33.5                  & \multicolumn{1}{c|}{14.4}                  & 38.1                  \\ \hline
\textbf{SCRNN}     & \multicolumn{1}{c|}{15.3}                  & 41.1         & \multicolumn{1}{c|}{14.8}         & 52.4         \\ \hline
\textbf{LSTM}      & \multicolumn{1}{c|}{18.5}                  & 27.6                  & \multicolumn{1}{c|}{10.0}                  & 21.7                  \\ \hline
\textbf{GRU}       & \multicolumn{1}{c|}{10.9}                  & 25.9                  & \multicolumn{1}{c|}{8.0}                   & 14.2                  \\ \hline
\end{tabular}}
\label{hdc-resiliency-improvement}
\end{table}

\textbf{HD Resiliency Improvement:} We calculate HD resiliency improvement over the selected DL models using Equation \ref{eq-improvement} for each STR configuration. Then, we find the maximum and average improvement for each DL method. Table \ref{hdc-resiliency-improvement} demonstrates the HD average and maximum resiliency improvement over the selected DL methods. For drive end experiment, HD improves DL model resiliency by up to 61.9\% where this number rises to 67.5\% for fan end dataset. Compared to the most resilient DL method (GRU), HD improves the resiliency by up to 25.9\% and 14.2\% for drive end and fan end data sets respectively. We are able to verify that HD provides a resilient learning solution against adversarial attacks.

\textbf{Training Overhead Comparison:} Table \ref{training-overhead-comparison} presents target models' training time (in seconds). In this table, each row represents a different target model (where the models are ordered in decreasing training overhead) and each column corresponds to the selected STR. We can observe that \textbf{HD} is the \textbf{most lightweight} model across all STRs. In the last column of this table, we share the average (across sample training ratios) normalized training time with respect to HD. HD can achieve up to 25.1$\times$ training speed up compared to DL methods. Compared to the most resilient DL method (GRU), HD brings 23$\times$ faster training. By this analysis, we can conclude that HD also provides a computationally efficient learning solution while being resilient to adversarial attacks. Training overhead is especially critical for I-IoT systems since data is collected continuously. When new data arrives, learning models require retraining to keep their prediction performances at a certain level \cite{gungor2021dowell}. HD can alleviate this retraining overhead due to its lightweight feature.

\begin{table}[]
\centering
\caption{Target Models Training Time Comparison}
\scalebox{0.8}{
\begin{tabular}{|c|cccccc|c|}
\hline
                & \multicolumn{6}{c|}{\textbf{Sample Training Ratio (\%)}}                                                                                                                                               & \textbf{Average}    \\ \hline
\textbf{Method} & \multicolumn{1}{c|}{\textbf{1.2}} & \multicolumn{1}{c|}{\textbf{2.4}}  & \multicolumn{1}{c|}{\textbf{4.8}}  & \multicolumn{1}{c|}{\textbf{9.6}}  & \multicolumn{1}{c|}{\textbf{19.4}} & \textbf{38.8}  & \textbf{Normalized} \\ \hline
\textbf{LSTM}   & \multicolumn{1}{c|}{151.0}        & \multicolumn{1}{c|}{366.1}         & \multicolumn{1}{c|}{514.8}         & \multicolumn{1}{c|}{980.4}         & \multicolumn{1}{c|}{1882.4}        & 3294.8         & 25.1                \\ \hline
\textbf{GRU}    & \multicolumn{1}{c|}{161.5}        & \multicolumn{1}{c|}{307.7}         & \multicolumn{1}{c|}{451.5}         & \multicolumn{1}{c|}{861.5}         & \multicolumn{1}{c|}{1623.3}        & 3165.5         & 23.0                \\ \hline
\textbf{CLSTM}  & \multicolumn{1}{c|}{16.0}         & \multicolumn{1}{c|}{31.8}          & \multicolumn{1}{c|}{81.0}          & \multicolumn{1}{c|}{174.6}         & \multicolumn{1}{c|}{413.7}         & 820.2          & 5.4                 \\ \hline
\textbf{CGRU}   & \multicolumn{1}{c|}{15.7}         & \multicolumn{1}{c|}{61.2}          & \multicolumn{1}{c|}{94.2}          & \multicolumn{1}{c|}{196.9}         & \multicolumn{1}{c|}{360.8}         & 727.5          & 5.1                 \\ \hline
\textbf{SCLSTM} & \multicolumn{1}{c|}{15.6}         & \multicolumn{1}{c|}{28.9}          & \multicolumn{1}{c|}{52.2}          & \multicolumn{1}{c|}{119.5}         & \multicolumn{1}{c|}{246.4}         & 488.6          & 3.3                 \\ \hline
\textbf{SCGRU}  & \multicolumn{1}{c|}{15.2}         & \multicolumn{1}{c|}{27.6}          & \multicolumn{1}{c|}{50.6}          & \multicolumn{1}{c|}{106.8}         & \multicolumn{1}{c|}{228.1}         & 448.5          & 3.1                 \\ \hline
\textbf{CRNN}   & \multicolumn{1}{c|}{10.8}         & \multicolumn{1}{c|}{19.4}          & \multicolumn{1}{c|}{36.1}          & \multicolumn{1}{c|}{77.4}          & \multicolumn{1}{c|}{167.6}         & 327.1          & 2.2                 \\ \hline
\textbf{SCRNN}  & \multicolumn{1}{c|}{7.7}          & \multicolumn{1}{c|}{13.8}          & \multicolumn{1}{c|}{26.9}          & \multicolumn{1}{c|}{53.6}          & \multicolumn{1}{c|}{98.6}          & 198.3          & 1.4                 \\ \hline
\textbf{HD}     & \multicolumn{1}{c|}{\textbf{6.0}} & \multicolumn{1}{c|}{\textbf{10.2}} & \multicolumn{1}{c|}{\textbf{18.6}} & \multicolumn{1}{c|}{\textbf{37.5}} & \multicolumn{1}{c|}{\textbf{72.0}} & \textbf{141.8} & \textbf{1.0}        \\ \hline
\end{tabular}}
\label{training-overhead-comparison}
\end{table}

\section{Conclusion}
\label{conclusion}
Industrial Internet of Things (I-IoT) is a notion that facilitates monitoring, automation and reliability of smart devices in production environments. I-IoT ensures that these devices are connected to the Internet which helps collecting big data, and utilizing this data to extract useful information. However, these interconnection brings numerous security vulnerabilities which can be exploited by an attacker to steal information or sabotage communication to harm the system. Attacks against machine learning (ML) is one type of cyberattack to deceive ML methods with fake inputs leading to worse prediction performance. Hyper-dimensional computing (HDC) is a novel learning solution which is robust against noise. In this paper, we utilize HDC to stay resilient against created black-box attack scenarios. Our experiments show that HDC can improve the resiliency of the state-of-the-art DL methods by up to 67.5\%. HDC can also achieve up to 25.1$\times$ training speed up compared to DL methods, providing a lightweight learning solution. This means that HDC can still perform well and efficiently under adversarial attacks which leads to more accurate replacement and maintenance decisions even under cyberattacks.         


\bibliographystyle{ieeetr}
\bibliography{bibfile}

\end{document}